This article appeared in PHYSICA B-CONDENSED MATTER Vol.71, 417256 (2025). DOI10.1016/j.physb.2025.417256.

# A phenomenological magnetomechanical model for hysteresis loops.

Alexej Perevertov

**AFFILIATION**

FZU - Institute of Physics of the Czech Academy of Sciences, Department of Magnetic Measurements and Materials, 18200 Prague, Czech Republic

Email: perever@fzu.cz

## ABSTRACT

In this work we propose a simple phenomenological model for magnetization curves of stressed samples. The effect of stress is introduced by scaling the arctangent function argument (magnetic field) proportionally to stress. The magnetization curve is modelled by one or two arctangent functions. Despite of its simplicity, the model gives a very good agreement with experimental curves, reproducing all stress-induced features usually observed on the magnetization curves including the common crossover point and constricted hysteresis loops. The popular effective field concept in this model is just a consequence of a simple scaling of the magnetic field. We show that the differential susceptibility is inversely proportional to the applied stress for all magnetization values. We propose to analyze the field and differential susceptibility as a function of magnetization in contrast to convenient *M(H)* loops.

## I. INTRODUCTION

Ferromagnetic materials like steels are widely used in industry – transformers, sensors, electronics etc. Their performance can be influenced by stress [1-9]. Mechanical stress can strongly affect the magnetization curve, *M(H)* curve of a ferromagnetic material. There are hundreds of works devoted to modeling the effect of stress on the magnetic susceptibility and hysteresis loops [10-29]. The comprehensive review of magneto-mechanical models can be found in the work of A. Kumar et al. [30].

One of the most popular modeling technique is the effective field by stress, which is added to the Jiles-Atherton model of hysteresis [31-32]. In this model the starting point is the anhysteretic curve. Then the hysteresis introduced by the pinning factor $k$ and the reversibility factor $c$ [31-32]. The anhysteretic *M(H)* curve is obtained by the solution of the equation $M = L(H+aM)$, where L is the Langevin function. The effective field idea was first proposed by Weiss [33], representing the internal coupling of the magnetic domains. Later Sablik et al. added the effective field by stress to the Jiles-Atherton model, which is the function of the stress and the magnetization [15]. A number of analytical functions of stress and magnetization were proposed [14-16,23].

The main problem of this approach is that it ignores the magnetic domain patterns by stress that are crucial for the shape of the magnetization curve. In the iron single crystal there are three magnetic easy axes. When a tensile stress is applied and the magnetization is measured along the stress direction, one of three axes closest to the stress direction is favored and far from the saturation the magnetization changes mostly by magnetic domains motion oriented along that axis [34].

In the case of compression one of three axes closest to the perpendicular to the stress direction is favored. Then at low fields the magnetic domains are oriented along this axis and at some critical field they suddenly reorient along the axis closest to the field direction [34-40]. This results in a so-called constricted or wasp-wasted curve with two-peak derivative [36]. Obviously, the anhysteretic curve of a compressed sample can't be represented by the Langevin function – it has to be also constricted. Recently double Langevin function was proposed for constricted loops of compressed or two-phase samples [37]. The situation in a polycrystalline iron or steel sample is even more complex – for applied compressive stress between 0 and 5 Mpa at low fields there is a mixture of normal domains along the field direction and the stress pattern with magnetization perpendicular to the applied field [34-35]. Only above 5 MPa the stress pattern at zero field finally occupies all the sample volume. For tensile stress there is the same several MPa transition period by a different reason – appearance of demagnetizing fields due to supplementary magnetic domains removal by stress [34]. So despite the fact that the tensile stress favors the magnetic domains in the field direction, the magnetization at low and moderate fields decreases with stress increase. From other side, at higher fields above the Villary reversal point the magnetization of iron or steel continuously increases with stress [1,9,34-35].

The effect of tensile stress on the magnetization curve of steel is much smaller comparing to compression, so the effect of compressive stress on the magnetization curve is much more important for investigation and modeling. More generally, the largest effect stress makes on a magnetization curve along the magnetically harder direction. For negative magnetostriction materials like nickel it is the tensile stress direction.

In our article on a plastically deformed steel we introduced an effective magnetic field by stress, $H_a$ as a product of a function of stress and a function of the magnetic field [36]:

$$H_a(\sigma,M) = g(\sigma)f(M) \tag{1}$$

It produces the coincident point (common crossover) phenomenon usually observed on hysteresis loops of stressed materials – all the curves cross at the same point between the coercive field and remanence [37-43]. This point is given by $f(M) = 0$.

This form of the effective field gave also a very good agreement between the experimental and calculated magnetization curves of a quenched and tempered spring steel [44]. In our works on Fe-3%Si steels under applied elastic stress we normalized the effective field by stress per one MPa. The magnetization at any stress in a given range ($\sigma_{MIN}$, $\sigma_{MAX}$) can be implicitly calculated as

$$H(\sigma,M) = H(\sigma_{MIN}, M) + k(\sigma)\,[H(\sigma_{MAX}, M) - H(\sigma_{MIN}, M)]/(\sigma_{MAX} - \sigma_{MIN}) =$$

$$= H(\sigma_{MIN}, M) + k(\sigma)\,\Delta H/\Delta\sigma(M) \qquad (2)$$

*where* $H(\sigma,M)$ - field in the material for a given stress, $\sigma$ and magnetization, $M$,

$\sigma_{MIN}$ - the minimum stress, from which above relation is valid,

$\Delta H/\Delta\sigma(M)$ - the effective field per one MPa,

$k(\sigma)$ - the function of stress only ( the stress coefficient).

We found that the function $k(\sigma)$ in the case of the applied elastic stress is a linear function of stress [34-35,37-40].

$$H(\sigma,M) = H(\sigma_{MIN}, M) + (k0 + c\sigma)\Delta H/\Delta\sigma(M) \qquad (3)$$

From other side, we have noticed that hysteresis loops of all steels we measured for the last 25 years follow the arctangent function including materials under stress [45]. So, from one side the ascending branches of quasistatic hysteresis loops (magnetization curves) of stressed steels can be described by introducing the effective field by stress, which is the product of a function of the stress and a function of the field, from other side the magnetization curve has a shape of one or two arctangent functions. In this work we propose a simple model that combine these two phenomena.

## II. APPLIED TENSION ALONG THE MAGNETIC FIELD DIRECTION

First, let us consider the case of the magnetization along the easy axis of a steel sample under tension. Iron and steels have the positive magnetostriction, so one expect a rectangular hysteresis loop due to stress. It is true up to a few MPa – the slope of the *M(H)* curve increases with stress, but then the opposite process starts. The reason is the removal of the supplementary domains by the stress that leads to demagnetizing fields at the sheet surface and grain boundaries [34]. The effect is much smaller comparing to compression and decreases with increase of the grain size and hence with decrease of the grains misorientation [34-41]. The *M(H)* curve of a steel under tension can be described by a single arctangent function:

$$M(H) = (2/\pi)\, M^{SAT} \arctan(a[H-H_C]) \tag{4}$$

where $M^{SAT}$ is the saturation magnetization, $H$ – internal magnetic field, $H_C$ – the coercive field, $2/\pi$ is the scaling coefficient since the maximum of the arctangent function is $\pi/2$.

In Fig.1 is the result of the fitting of the *M(H)* curve by the least-square method for the Goss-textured Fe-3%Si steel (the small grains sample in [34]) under tensile stress from 9 to 97 MPa by the arctangent function:

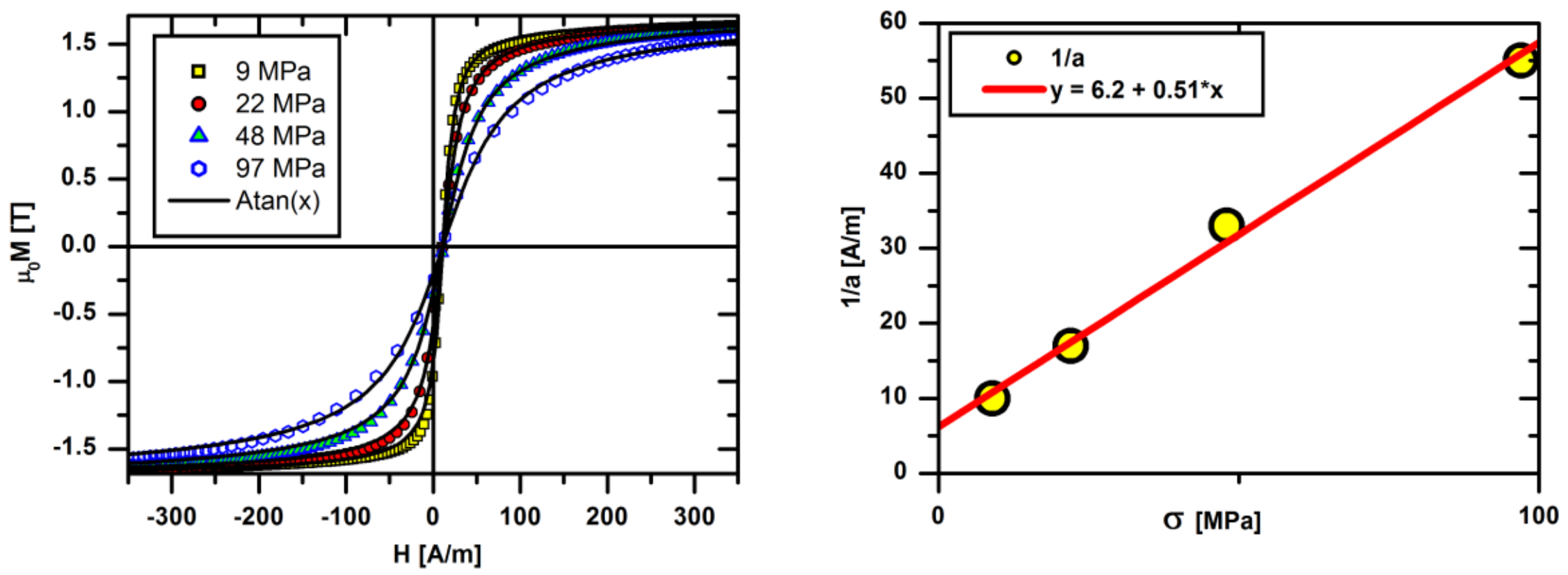

(a) (b)

**Fig.1.** Fit of the ascending branches of hysteresis loops (*M(H)* curves) of a Goss-textured Fe-3%Si steel under tension (a) and the stress dependence of the parameter, *a* (b). Other parameters are constant: $\mu_0 M^{SAT} = 1.85T, H_C = 11\ A/m.$

Here the coefficients $M^{SAT}$ and $H_C$ were the same for all stresses. The only parameter that changes with stress is the curves slope, *a*.

So, we can further simplify the Eq. (4):

$$1/a = b_0 + b_1\sigma;\ H_C = const;\ M^{SAT} = const; \tag{5}$$

The magnetic field as a function of stress and magnetization is given by

$$tan([M/M^{SAT}]\,\pi/2)= (H\text{-}H_C)/\,(b_0 + b_1\sigma);$$

$$H(\sigma_2) - H(\sigma_1) = b_1(\sigma_2 - \sigma_1)\, tan([M/M^{SAT}]\,\pi/2); \qquad (6)$$

The function of magnetization in the effective field is a simple tangent function. Experimental effective field follows the tangent function up to some magnetization value and then it decreases at higher field and becomes negative [18,34-35,39] . The reason is a small change in the maximum magnetization. If the maximum magnetization slightly increases with the tensile stress, the effective field at higher fields deviates from the tangent function (see Fig.2). Such a shape of the effective field have been observed before on a number of stressed steels [18,34-35]. The slightest increase in the maximum magnetization, 1% or even less, produces such a shape. It leads to a well-known effect – the Villary reversal point, the field value after which the magnetization increases with stress [1,34-35].

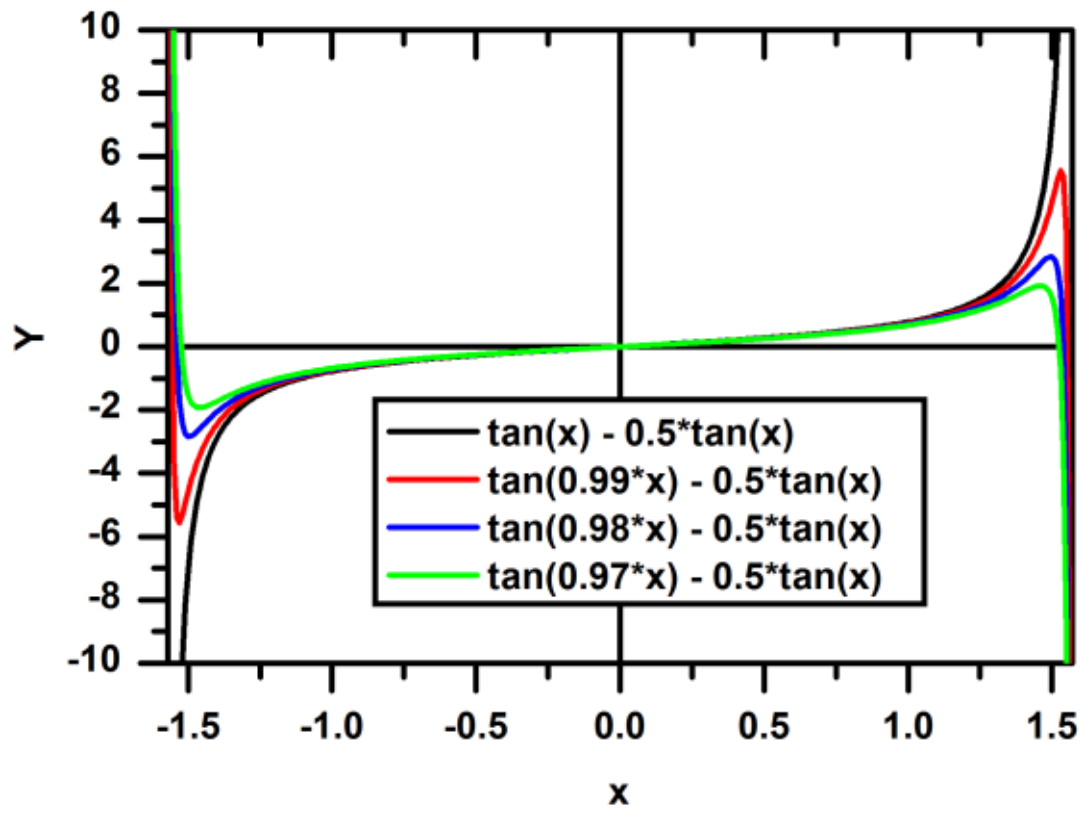

**Fig.2.** The effective stress field for a maximum magnetization increase with stress by 0, 1, 2 and 3%.

## III. APPLIED COMPRESSION ALONG THE MAGNETIC FIELD DIRECTION

Now let us consider the effect of compression, which has much higher effect on magnetization curves comparing to the tension. Under compression the hysteresis loop of a steel becomes constricted (has a butterfly or wasp-wasted shape) – it looks like a sum of two hysteresis loops. We can write the magnetization as a sum of two arctangent functions from (4):

$$M(H) = (2/\pi)\, M^{SAT}\, (arctan(a_1[H+H_{C1}]) + arctan(a_2[H-H_{C2}])); \qquad (7)$$

where $a_i$ *and* $H_{Ci}$ – are corresponding parameters of each half and the amplitude is the same.

In Fig.3 experimental *M(H)* curves of a Goss-textured Fe-3%Si sheet [38-39] and fits by a sum of two arctangent functions are shown.

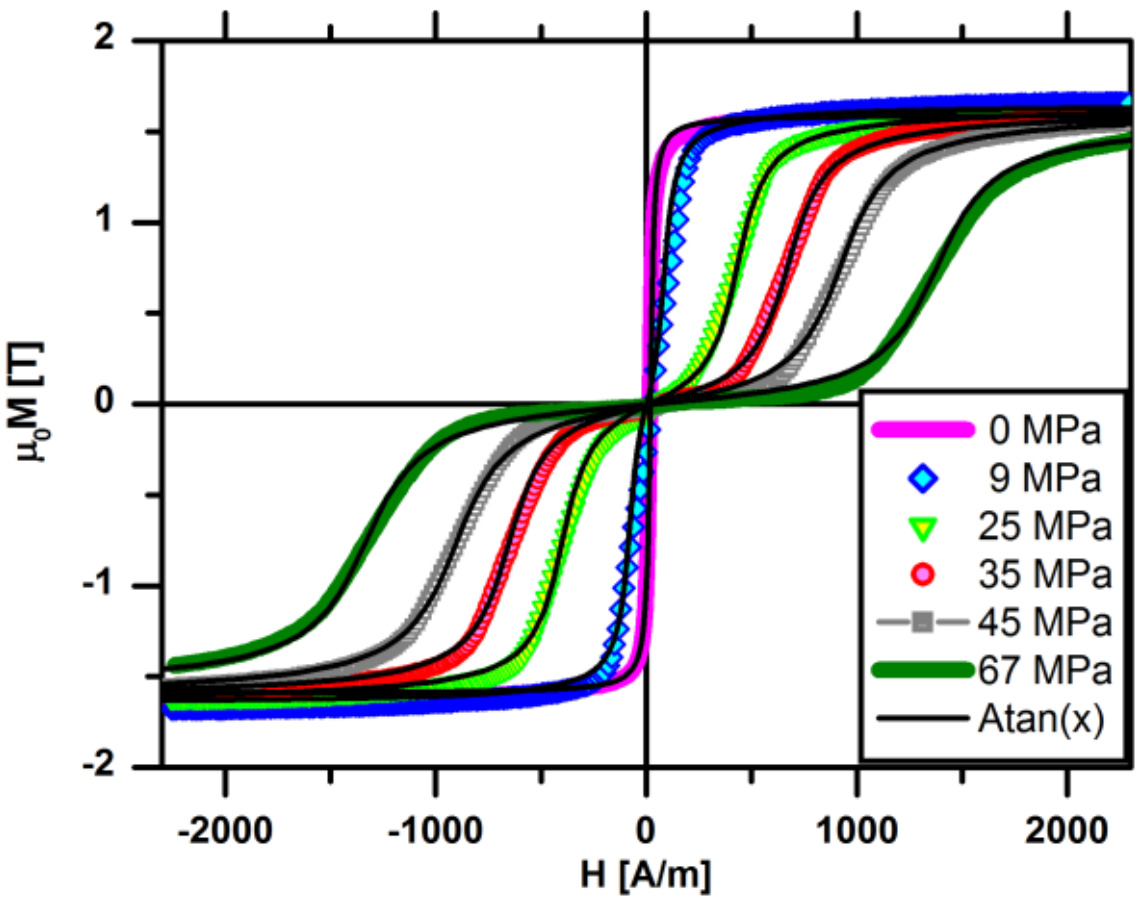

**Fig.3.** *M(H)* curves of a Goss-textured Fe-3%Si sheet under compression and corresponding arctangent fits.

The parameters of arctangent functions are given in the Table I and in Fig.4.

Table I. The parameters of arctangent functions for *M(H)* curves of a Goss-textured Fe-3%Si sheet under compression.

| $\sigma$, *MPa* | $(2/\pi)\ M^{SAT}$, $10^6$ *A/m* | *a1*, *m/A* | *a2*, *m/A* | $H_{C1}$, *A/m* | $H_{C2}$, *A/m* |
|---|---|---|---|---|---|
| 0 | 1.015 | 0.089 | -- | 17.5 | -- |
| 9 | 0.525 | 0.025 | 0.025 | 80 | 82 |
| 25 | 0.525 | 0.01 | 0.01 | 400 | 432 |
| 35 | 0.52 | 0.0075 | 0.0075 | 163 | 168 |
| 45 | 0.52 | 0.0055 | 0.006 | 904 | 924 |
| 67 | 0.506 | 0.0045 | 0.005 | 1332 | 1380 |

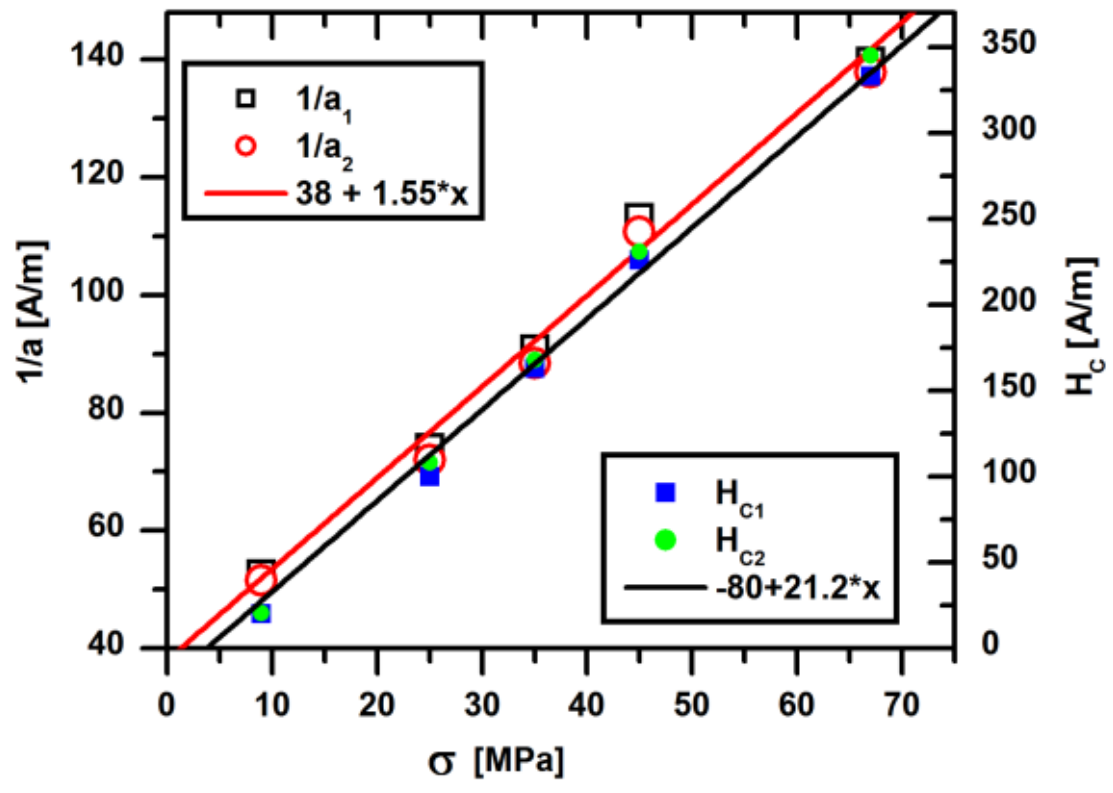

**Fig.4.** The parameters of arctangent functions from Eq.(7) for a compressed Goss-textured Fe-3%Si sheet as a function of stress.

The widths (~1/a ) and the positions of two arctangent functions change linearly with stress. The slope parameters, $a_1$ and $a_2$ are equal and the difference in positions $H_{C2}$ and $H_{C1}$ should give the total coercive field, $H_C$.

$$a_1 = a_2 = 1/(k_0 + k\sigma);\ H_{C2} = H_{C1} + 0.5H_C = l_0 + l\sigma \tag{8}$$

The value of the critical field can be estimated by considering an ideally oriented crystal with no demagnetizing fields considering the energy balance as [35,40]:

$$H_{CR} = (3/2)\lambda_{100}\sigma/Ms \tag{9}$$

where $\lambda_{100}$ is the magnetostriction constant along the [100] direction, $\sigma$ is the compressive stress, and $M_S$ is the saturation magnetization.

We see that not only critical fields change proportionally to stress but also the magnetization curve width.

## IV. RESIDUAL STRESS, PLASTIC DEFORMATION.

Above we considered materials under an applied elastic stress. All *M(H)* curves crossed at the coercivity and the arctangent parameters changed linearly with stress. In the case of plastically deformed samples in the condition of the residual stress the coincidence point usually lies between the coercivity and the remanence [36-43]. To simulate *M(H)* curves of plastically deformed sample with the coincident point at the arbitrary field $H_{COIN}$ the equation for a magnetization curve can be written in the next form:

$$M(H) = (2/\pi)M^{SAT}(\arctan(a(H-H_{COIN})+b_1)+\arctan(a(H-H_{COIN})+b_2));\ b_1=const;\ b_2=const; \qquad (10)$$

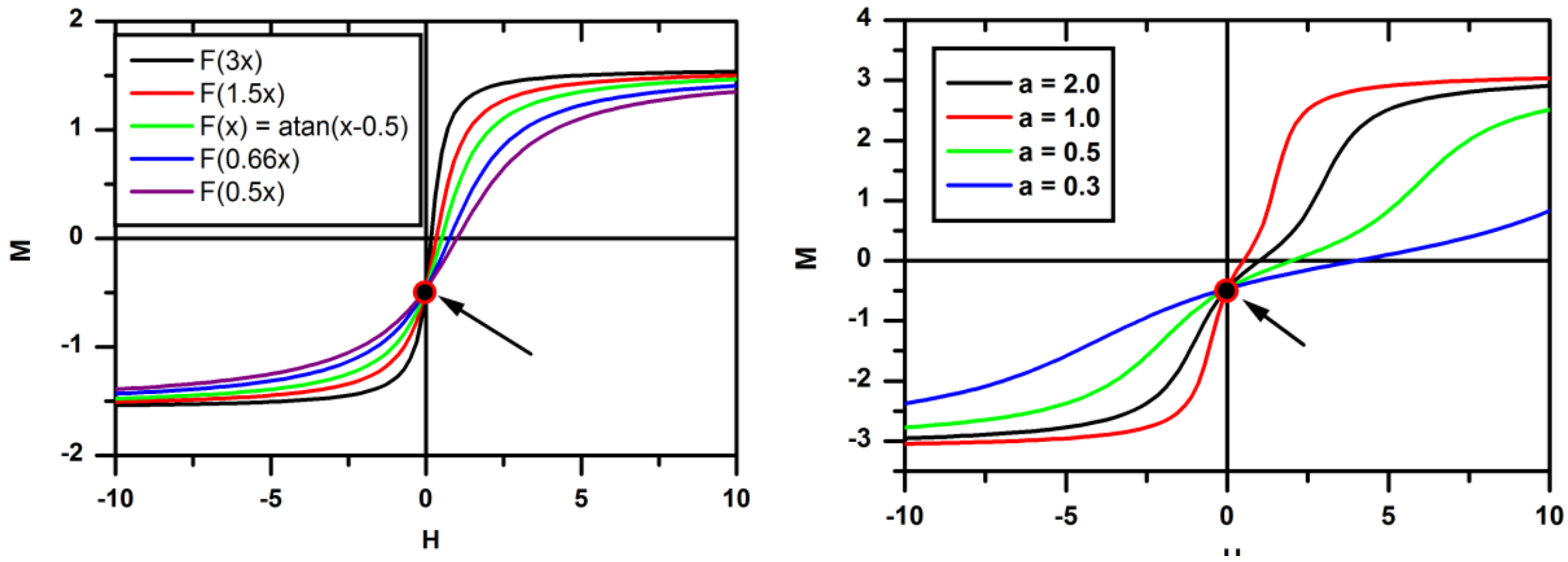

(a) (b)

**Fig.5** *M(H)* curves with the coincident point at the remanence for a magnetization along the easy magnetization direction (a) and along the hard one (b).

In Fig.5 the *M(H)* curves are shown for $H_{COIN}$ = 0 (remanence) are shown. Since all experimental data so far report the coincidence point between the remanence and the coercive field, $H_C$, so $0<H_{COIN}<H_C$.

## V. DISCUSSION

In 1930 Kersten analyzed the total energy of a stressed ferromagnet and concluded that the magnetization slope, the relative susceptibility should be inversely proportional to the stress for the initial part of the magnetization curve [3]:

$$\chi = M/H = J_S^2/(3\mu_0\gamma_S\sigma) \qquad (11)$$

where $\chi$ is the relative susceptibility; $H$ – the magnetic field; $M$ – magnetization; $J_S$ – the saturation magnetic polarization; $\gamma_S$ - the saturation magnetostriction and $\sigma$ is the applied stress.

This formula was obtained by assuming that the magnetocrystalline anisotropy is much smaller comparing to stress-induced one and can be neglected. The *M(H)* curve in this approximation is just a straight line and the magnetization changes by magnetic moments rotation. In iron and iron-based steels it is just the opposite - the magnetocrystalline anisotropy is much larger and the magnetization changes by magnetic domain walls motion in each grain. However in this work we show that the the differential susceptibility of nonlinear *M(H)* curves of stressed polycrystalline iron-based ferromagnets is also inversely proportional to stress for any magnetization value (not the field). We propose to plot and analyze the differential susceptibility and the field as function of the magnetization in contrast to generally accepted *M(H)* and $\chi(H)$ curves since it simplifies the analysis of the curves – the field at any magnetization value is proportional to the stress and the differential susceptibility is inversely proportional to it. It can be seen in Fig.6 that the differential susceptibility as a function of the field changes peaks positions and widths with stress, while when plotted as a function of the magnetization peaks coincide. The third central peak for 9 MPa either because it is still the transition or due to small bending that is unavoidable for compressed thin and long sheets.

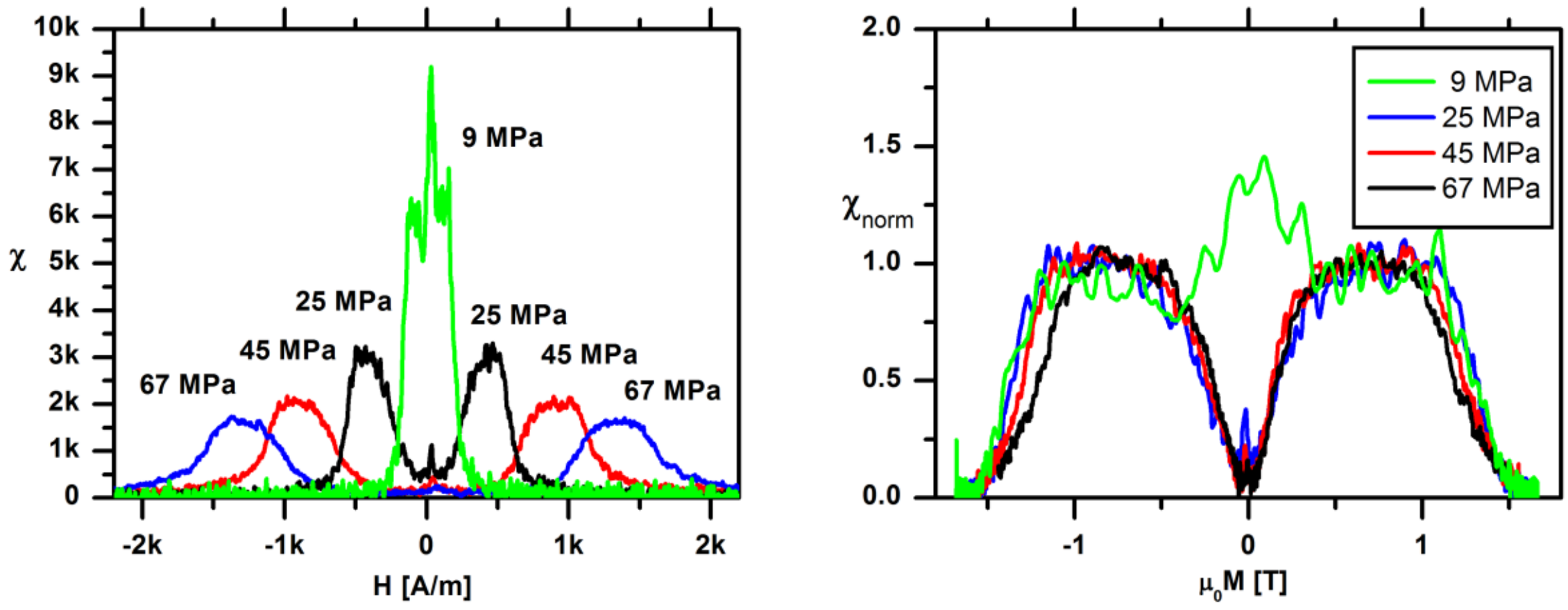

(a) (b)

**Fig.6.** $\chi$ *(H) (a)* and $\chi$*(M)* plots for compressed grain-oriented Fe-3%Si steel (see Fig.3).

In $\chi$*(H)* plots the peaks change their position and width [44], while peaks in $\chi$*(M)* plots have the same position and width with only one parameter changing with the stress – the vertical scaling (see Fig.6b).

It should be emphasized that if the effect of stress is just a scaling of the argument (magnetic field) in the magnetization function *M(H)* = *f(x)*, then the effective field concept is valid for any sigmoid function, not necessarily the arctangent function.

**SUMMARY.** To summarize, we propose a simple phenomenological model for magnetization curves of stressed samples. The complex effect of stress is reduced to a simple scaling of the arctangent function argument. The magnetization curve of stressed samples is modelled by one or two arctangent functions. For the case of applied elastic stress the scaling is linear with stress. The relation between the scaling coefficient and residual stress value should be investigated in future works. The popular concept of the effective field by stress is the direct consequence of the scaling of the magnetization function argument, the magnetic field of the zero-stress *M(H)* curve.

The model gives a very good agreement with experimental curves of stressed steel samples for stresses above a few MPa. We show that the Kersten relation for a relative susceptibility is also valid for the differential susceptibility of nonlinear *M(H)* curves of stressed polycrystalline iron-based ferromagnets for any magnetization value. We propose to plot and analyze the differential susceptibility and magnetization as function of the magnetization in contrast to generally accepted *M(H)* and $\chi(H)$ curves

## DATA AVAILABILITY

The data that support the findings of this study are available from the corresponding authors upon request. The data are also available in Zenodo at https://doi.org/10.5281/zenodo.15209969 [45].

## ACKNOWLEDGMENTS

The authors acknowledge the assistance provided by the Ferroic Multifunctionalities project, supported by the Ministry of Education, Youth, and Sports of the Czech Republic. Project No. CZ.02.01.01/00/22_008/0004591, co-funded by the European Union.